\documentclass{ws-ijmpd}
\usepackage[super,compress]{cite}

\def\tu{{\bar u}}
\def\tv{{\bar v}}
\def\tx{{\bar x}}

\def\trho{{\bar \rho}}

\begin{document}


%
\catchline{}{}{}{}{}
%

\title{Radiation from a D-dimensional collision of shock waves: \\  Two dimensional reduction and Carter-Penrose diagram}
\author{FL\'AVIO S. COELHO\footnote{flaviodscoelho@gmail.com}\vspace{1mm}}

\author{MARCO O. P. SAMPAIO\footnote{msampaio@ua.pt}\vspace{1mm}}

\address{Departamento de F\'isica da Universidade de Aveiro and\\ Center for Research and Development in Mathematics and Applications (CIDMA),\\
 Campus de Santiago, 3810-183 Aveiro, Portugal}

\maketitle


\begin{abstract}
We analyse the causal structure of the two dimensional (2D) reduced background used in the perturbative treatment of a head-on collision of two $D$-dimensional Aichelburg-Sexl gravitational shock waves. After defining all causal boundaries, namely the future light-cone of the collision and the past light-cone of a future observer, we obtain characteristic coordinates using two independent methods. The first is a geometrical construction of the null rays which define the various light cones, using a parametric representation. The second is a transformation of the 2D reduced wave operator for the problem into a hyperbolic form. The characteristic coordinates are then compactified allowing us to represent all causal light rays in a conformal Carter-Penrose diagram. Our construction holds to all orders in perturbation theory. In particular, we can easily identify the singularities of the source functions and of the Green's functions appearing in the perturbative expansion, at each order, which is crucial for a successful numerical evaluation of any higher order corrections using this method. 

\end{abstract}


\ccode{PACS numbers: 04.50.-h, 04.20.Cv, 04.30.Db, 04.50.Gh}

\tableofcontents

\section{Introduction}	

The problem of solving the Einstein field equations in the strong field regime has been, from the early days of General Relativity (GR), a very challenging one. In particular, exact analytic solutions in the highly dynamical regime are almost impossible to obtain, in realistic situations, and one has to resort to elaborate, and computationally intensive, numerical techniques. The numerical relativity community has made a great effort, in recent years, to perfect numerical techniques and to develop the tools necessary to produce numerical waveforms to describe astrophysical candidate events -- see Refs.~\cite{Berti:2015itd,Cardoso:2014uka} for a review. In parallel, the experimental community has achieved outstanding technological advances in the interferometry needed to directly detect such events and has finally provided us with a breakthrough: the first direct observation of one of the most fundamental predictions of GR -- the gravitational wave signal from a black hole binary merger~\cite{Abbott:2016blz} by the LIGO and Virgo collaborations.

Even though for highly dynamical processes, in GR, we are typically forced to use fully numerical strategies, there are particular limits where one may hope to get some insights on the underlying physics using semi-analytic techniques. Such is the case of a head on collision of two Schwarzschild black holes in the limit where both travel at the speed of light. In this limit, the system becomes equivalent to two colliding particles travelling precisely at the speed of light, each described by an Aichelburg-Sexl gravitational shock wave~\cite{Aichelburg:1970dh}. To solve this problem, a perturbative method was first developed in $D=4$ space-time dimensions by D'Eath and Chapman, and later by D'Eath and Payne~\cite{D'Eath:1992hb,D'Eath:1992hd,D'Eath:1992qu}. In recent years, the study of this problem found a renewed interest in the context of higher-dimensional brane-world models~\cite{ArkaniHamed:1998rs,Antoniadis:1998ig,Randall:1999ee,Randall:1999vf} with TeV scale gravity. In such a framework, it has been suggested that microscopic black holes (BHs) could be formed~\cite{Argyres:1998qn,Banks:1999gd} in realistic particle accelerators, such as the LHC~\cite{Giddings:2001bu,Dimopoulos:2001hw}, or in ultra-high energy cosmic ray collisions~\cite{Feng:2001ib,Anchordoqui:2001ei,Emparan:2001kf}. In fact, it has been argued earlier by 't~Hooft that the collision of two point-like particles at trans-Planckian center of mass energies should be well described by General Relativity~\cite{'tHooft:1987rb}. So far, no signs of TeV gravity have been found at the LHC. Nevertheless, it is still of general interest to place the best bounds on the fundamental Planck scale of such models~\cite{Chatrchyan:2012taa,Aad:2012ic,Chatrchyan:2013xva,Aad:2013gma,Aad:2015mzg} through a better phenomenological description~\cite{Cardoso:2012qm}.

Another approach to this problem consists of finding \textit{bounds} on the amount of gravitational radiation emitted (inelasticity) in these collisions which, in turn, places a bound on the production cross section for BH formation. This is achieved by studying the formation of trapped surfaces in the collision space-time of two Aichelburg-Sexl shock waves -- a method, originally due to Penrose in four space-time dimensions, that has been extended to $D$ dimensions and to other different situations~\cite{Giddings:2001bu,Yoshino:2002br,Yoshino:2002tx,Yoshino:2005hi,AlvarezGaume:2008fx,Albacete:2009ji}. However, there is an important difference between these calculations and the perturbative method of D'Eath and Payne. The latter is aimed at computing the metric in the future light-cone of the collision and to obtain \textit{estimates}, rather than bounds on the elasticity. Furthermore, this method has two virtues which encourages its study in $D>4$, namely: (i) at second order in perturbation theory the initial data for the pre-collision exact solution of the Einstein equations is fully taken into account; (ii) the corresponding result obtained by D'Eath and Payne, in $D=4$, for the inelasticity $\epsilon_{\rm 2nd \ order}=0.164$, agrees well with the extrapolation to the speed of light of results from numerical simulations of colliding BHs~\cite{Sperhake:2008ga} and other compact objects~\cite{East:2012mb,Rezzolla:2012nr}.

In previous papers~\cite{Herdeiro:2011ck,Coelho:2012sya,Coelho:2012sy,Coelho:2014gma}, we have extended several of the results of the method by D'Eath and Payne to $D$ dimensions. In the process, we have found a new strikingly simple formula for the amount of radiation emitted in the leading order approximation~\cite{Coelho:2012sya}. This we first found numerically and later we proved it to be exact~\cite{Coelho:2014gma}. Furthermore we have proved, for the first time, a conjecture which was implicit in the original calculations of D'Eath and Payne\cite{D'Eath:1992hb,D'Eath:1992hd,D'Eath:1992qu,DEath:1990de}: there is an exact correspondence between the order of the angular expansion of the inelasticity around the axis of symmetry and the order of the perturbative expansion. Additionally, we have found closed form analytic solutions for all surface terms contributing to the gravitational wave form which were also unknown even in $D=4$. Both of these results are related to a reduction of the problem to two dimensions, which we have extended for all $D$. 

In this article, we complete the two dimensional (2D) reduction of the problem by finding characteristic coordinates using two methods. First we use causality considerations to define the future light cone of the collision event and to define the past light cone of an observation event to the future of the collision. This is then used to construct parametric solutions for the curves describing the 2D reduced light rays of the various light cones, as well as the light rays corresponding to singularities of the gravitational source terms and of the Green's function. The second (independent) method consists of performing a hyperbolicity analysis where the wave operator, appearing at each order in the perturbative expansion, is reduced to a standard characteristic form with only (second order) mixed derivative terms. Finally, we compactify the characteristic coordinates and find a Penrose diagram which clarifies the causal structure of the 2D reduced space-time and sets the stage for all higher order calculations. 

The structure of the paper is the following. In Sect.~\ref{sec:overview} we start with a brief review of the setup for the problem. In Sect.~\ref{red2D} we analyse in detail the causal structure of the problem in the 2D reduced description, by defining the future light cone of the collision (Sect.~\ref{future_cone}) and the past light cone of an observation event (Sect.~\ref{past_cone}). After identifying the characteristic coordinates, we build them again with an independent method in Sect.~\ref{characteristics}. In Sect.~\ref{CP_diagram} we discuss the conformal diagram for the problem and the asymptotics for the metric functions and for the Green's functions, at any order in perturbation theory, for an observer at null infinity. Our conclusions are summarised in Sect.~\ref{sec:conclusions}.

\section{Overview of the perturbative method and previous results}
\label{sec:overview}
In this section, we provide a brief summary of the perturbative method. Complete details can be found in Refs.~\cite{Herdeiro:2011ck,Coelho:2012sya,Coelho:2012sy,Coelho:2014gma,Coelho:2014qla,Coelho:2013zs,Coelho:2014cna} and in the early papers by D'Eath and Payne~\cite{D'Eath:1992hb,D'Eath:1992hd,D'Eath:1992qu,DEath:1990de}. Pedagogical reviews can also be found in Refs.~\cite{Sampaio:2013faa,Coelho:2015qaa}.

\begin{figure}[t]
\begin{center}
\includegraphics[width=1\linewidth]{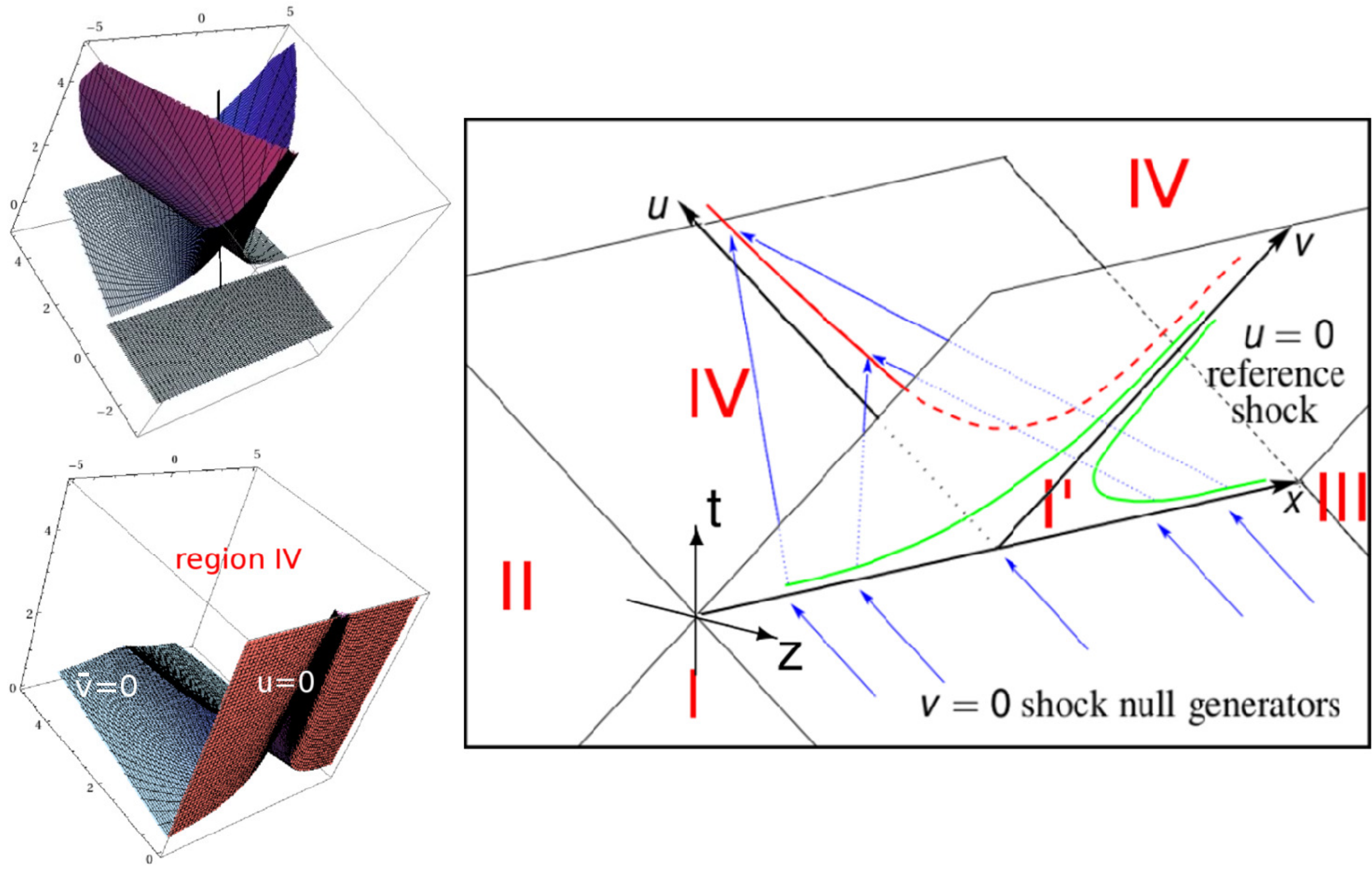}
\end{center}
\caption{\label{5DspacetimeDiag}\emph{Right:} 3D space-time diagram which shows the $(t,z,\rho\equiv x)$ axes, four regions (in red roman numerals), the generators of the $v$-shock (in blue) and the collision surface (in green). \emph{Adapted from Ref.~\cite{Herdeiro:2011ck}.} \emph{Left:} The top diagram shows a surface defined by the $v=0$ shock null generators (when $u<0$) as they scatter through the $u=0$ shock (when $u>0$). The bottom one shows the union of the two causal surfaces ($u=0\wedge v>0$ and $\bar{v}=0\wedge u>0$) which define the future light cone of the collision (region IV). \emph{Adapted from Ref.~\cite{Sampaio:2013faa}.}}
\end{figure}

The basic premiss of this method comes from the observation that two light-like particles, i.e. travelling at the speed of light, can not influence each other before they collide. Thus, if one knows the gravitational field of one such particle moving along the direction of the positive $+z$ axis, we can superpose another similar solution moving in the opposite direction. The line element for the superposition is then exact outside the future light cone of the collision, and one can show that it can be conveniently written in an asymmetric system of coordinates which is adapted to one of the particles (say the particle moving along $u=t-z=0$)~\cite{Herdeiro:2011ck,Sampaio:2013faa,Coelho:2015qaa}:
\begin{multline}ds^2 = -du dv +\delta_{ij}dx^idx^j+\kappa \Phi(\rho) \delta(u) du^2+\\ \left\{-2\bar{h}(u,v,\rho)\bar{\Delta}_{ij}+\bar{h}(u,v,\rho)^2\left((D-3)\delta_{ij}-(D-4)\bar{\Delta}_{ij}\right)\right\}d\tx^id\tx^j 
 \ .\label{collisionBrink}
\end{multline} The coordinates $u,v$ are retarded and advance null coordinates, $t$ is a time coordinate and $z$ a coordinate along the axis of symmetry. The coordinates $x^i$ are Euclidean coordinates on the plane transverse to the collision axis (the $z$-axis). The first line of Eq.~\eqref{collisionBrink} corresponds to the Aichelburg-Sexl solution for the reference point particle, $u=0$, and contains a delta-function like impulsive part with a $\rho\equiv x^ix_i$ dependent profile which is~\cite{Eardley:2002re}
\begin{equation}
\Phi(\rho)=\left\{
\begin{array}{ll}
 -2\ln(\rho)\ , &  D=4\  \vspace{2mm}\\
\displaystyle{ \frac{2}{(D-4)\rho^{D-4}}}\ , & D>4\ \label{phidef}
\end{array} \right. \ .
\end{equation}
The coordinate $v=t+z$ is such that the left moving particle, colliding head on, travels with $v=0$. Here $\kappa\equiv 8\pi G_D E/\Omega_{D-3}$ is the only parameter in the problem, corresponding to the energy parameter of each of the colliding point like particles. $E$ is the energy of the point-like particle, $G_D$ the $D$-dimensional Newton's constant and $\Omega_{D-3}$ is the area of the $D-3$ sphere. Observe that the line element on the first line of Eq.\eqref{collisionBrink} is flat space in the usual Minkowski coordinates except for the last impulsive term. In the second line of Eq.~\eqref{collisionBrink} we have that 
\begin{eqnarray}
\bar{h}(u,v,\rho)&\equiv&-\dfrac{\kappa\Phi'\tv}{2\trho}\theta(\tv)\\
\bar{\Delta}_{ij}&\equiv& \delta_{ij}-(D-2)\bar{\Gamma}_i\bar{\Gamma}_j\\
\bar{\Gamma}_i&\equiv& \frac{\bar{x}_i}{\bar{\rho}}
\end{eqnarray}
where the barred coordinates are related to the un-barred ones through
\begin{eqnarray}
u &=& \tu\ ,\nonumber \\
v &=& \tv+\kappa\theta(\tu)\left(\Phi + \frac{\kappa \tu (\bar{\nabla}\Phi)^2}{4}\right)\ = \tv+\kappa\theta(\tu)\left(\Phi + \frac{\kappa \tu \Phi'^2}{4}\right),\nonumber \\
x^i&=& \tx^i + \kappa\frac{\tu}{2} \bar{\nabla}_i \Phi(\tx)\theta(\tu) \Rightarrow \left\{\begin{array}{rcl} \rho&=& \trho\Big(1+ \frac{\kappa \tu \, \theta(\tu)}{2 \bar \rho}\Phi'\Big)  \\
\phi_{a}&=&\bar \phi_a  \end{array}\right.
\ . \label{ct}
\end{eqnarray}
In a region $\bar{h}\ll 1$, the terms on the second line of Eq.~\eqref{collisionBrink} can be viewed as a perturbation due to the $v=0$ particle. In such a region we can approximate the solution to the Einstein equations perturbatively around the reference $u=0$ shock wave~\cite{Sampaio:2013faa}. In fact, the exact metric in the boundary region separating the future light cone of the collision (where the line element is unknown) from the past of the collision (where the superposition is exact) is given by the following two branches:
\begin{equation}
g_{\mu\nu}(u>0,\tv=0,x_i)=\eta_{\mu\nu} \; ,\label{InitCond1}
\end{equation}
 and
\begin{equation}
g_{\mu\nu}(u=0^+,v,x_i)\equiv\eta_{\mu\nu}+h_{\mu\nu}=\eta_{\mu\nu}+h_{\mu\nu}^{(1)}+h_{\mu\nu}^{(2)}\;.\label{InitCond2}
\end{equation}
Here the $h_{\mu\nu}^{(i)}$ can be obtained from~\eqref{collisionBrink} and the label $i=1,2$ corresponds to the power of $\bar{h}$ appearing in each term. The two surfaces where these conditions are set can be better visualised in the diagram of Fig.~\ref{5DspacetimeDiag} (left bottom surface).

The perturbative method, consists of assuming a perturbative ansatz in the future region of the collision. The main steps are:
\begin{enumerate}
\item Assume a perturbative ansatz with the form (here $k$ can be identified as the order in an expansion in $\kappa^k$):
\begin{equation} 
g_{\mu\nu}=\eta_{ \mu \nu}+\sum_{i=1}^\infty h_{\mu\nu}^{(k)}\ ,
\label{eq:pertexpansion}  
\end{equation}
\item Insert the ansatz in the Einstein equations, equate order by order, impose the de Donder gauge condition $\bar{h}^{(k)\alpha\beta}_{\phantom{(n)\alpha\beta},\beta}=0$ on the trace reversed metric perturbation $\bar{h}^{(k)\alpha\beta}$, and obtain a tower of wave equations
\begin{equation}\label{eq:nth_order_Feq}
\Box h^{(k)}_{\mu\nu}=T^{(k-1)}_{\mu\nu}\left[h_{\alpha\beta}^{(m<k)}\right] \ .
\end{equation}
Here the right hand side source at order $k$ depends only on metric perturbations of lower order, $h^{(k)\alpha \beta}$ are the metric perturbations in de~Donder coordinates, and $\Box=-2\partial_u\partial_v+\partial_i\partial^i$ is the usual wave operator in Minkowski space. 
\item Since the solution of the wave equation with sources is well known, the general formal solution, at each order $k$, is
\begin{equation} \label{eq:sol_orderbyorder}
h^{(k)}_{\mu\nu}(y)=F.P.\int_{u'>0}d^{D}y'\, G(y,y')\left[2\delta(u')\partial_{v'}h^{(n)}_{\mu\nu}(y')+T^{(k-1)}_{\mu\nu}(y')\right] \ ,
\end{equation}
where $G(y,y')$ is the retarded Green's function which propagates the source from the point $y'$ inside the past light cone of the observer, to the observation point $y$, and $F.P.$ denotes the finite part of the integral~\cite{Herdeiro:2011ck,Sampaio:2013faa,Coelho:2015qaa}.
\end{enumerate}
Thus the problem reduces to evaluating the integrals Eq.~\eqref{eq:sol_orderbyorder} order by order. The problem simplifies even further by using the axial symmetry to reduce the metric perturbations and the source tensors. The general reduction we have found in $D$ dimensions is as follows~\cite{Coelho:2012sy,Sampaio:2013faa,Coelho:2014gma}:
\begin{eqnarray}
&h_{uu}\equiv A=A^{(1)}+A^{(2)}+\ldots \qquad &h_{ui}\equiv B \,\Gamma_i =(B^{(1)}+B^{(2)}+\ldots)\Gamma_i  \nonumber\\
&h_{uv}\equiv C=C^{(1)}+C^{(2)}+\ldots \qquad &h_{vi}\equiv F \,\Gamma_i =(F^{(1)}+F^{(2)}+\ldots)\Gamma_i  \label{app:gen_perts}\\
&h_{vv}\equiv G=G^{(1)}+G^{(2)}+\ldots \qquad &h_{ij}\equiv E \,\Delta_{ij}+H\, \delta_{ij} = (E^{(1)}+\ldots) \Delta_{ij}\nonumber \\
& &\phantom{h_{ij}\equiv E \,\Delta_{ij}+H\, \delta_{ij} =}+(H^{(1)}+\ldots) \delta_{ij} \ .\qquad\nonumber 
\end{eqnarray}
Where all the functions $A^{(k)},B^{(k)},\ldots$ depend only on the coordinates $(u,v,\rho)$. Similarly, for the sources $T^{(k-1)}_{\mu\nu}$, we can find such a decomposition.  The radiative components are for example
\begin{eqnarray}
T^{(k-1)}_{ij}&=&T_H^{(k-1)}(u,v,\rho)\delta_{ij}+T^{(k-1)}_E(u,v,\rho)\Delta_{ij}\,.
\end{eqnarray}
In the remainder we denote the functions in~\eqref{app:gen_perts} collectively by $F^{(k)}(u,v,\rho)$. The associated source, \textit{i.e.} the projection of $T^{(k-1)}_{\mu\nu}$ with the same basis tensor, is denoted by $T^{(k-1)}(u,v,\rho)$. Then from~\eqref{eq:sol_orderbyorder} one obtains
\begin{multline}
F^{(k)}(u,v,\rho)=F.P.\int_{u'>0} d^{D}y'\, G(y,y')\Lambda_m\left(\frac{x\cdot x'}{\rho\rho'}\right)\left[T^{(k-1)}(u',v',\rho')+\right. \\
\left.\phantom{T^{(k-1)}}2\delta(u')\partial_{v'}F^{(k)}(0,v',\rho')\right]\,,\label{general_F}
\end{multline}
where the projection scalars are (for $m=\{0,1,2\}$),
\begin{equation}
\Lambda_m(z)\equiv \left\{1,z,(D-3)^{-1}\left((D-2)z^2-1\right)\right\}\,.
\label{lm}
\end{equation}
Here $m$ is the rank of the axial basis tensor corresponding to $F$ (and similarly for $T$). In particular, $m=0$ for $A,C,G,H$, $m=1$ for $B,F$ and $m=2$ for $E$ -- see Eq.~\eqref{app:gen_perts}. For notational simplicity, in the remainder, we denote the sum of both volume and surface sources by:
\begin{equation}
S^{(k)}(u,v,\rho)\equiv T^{(k-1)}(u,v,\rho)+2\delta(u)\partial_{v}F^{(k)}(0,v,\rho) \ .
\end{equation}
Finally, observe that the Green function only depends on the quantity 
\[\chi\equiv-\eta_{\mu\nu}(y-y')^\mu (y-y')^\nu\; .\]

\section{The two-dimensionally reduced problem}
\label{red2D}
In Ref.~\cite{Coelho:2014gma} we have shown that there is a generalisation to $D>4$ of the conformal symmetry found in Ref.~\cite{D'Eath:1992hd} which allowed a further separation of variables in $D=4$. This symmetry implies that the $\rho$ dependence can be completely factored out, and the problem becomes effectively two-dimensional. Both the metric functions and respective sources can then be decomposed as
\begin{equation}
F^{(k)}(u,v,\rho)= \dfrac{f^{(k)}(p,q)}{\rho^{(D-3)(2k+N_u-N_v)}}\,,\label{f_pq}
\end{equation}
and
\begin{equation}
S^{(k)}(u,v,\rho)=\dfrac{s^{(k)}(p,q)}{\rho^{(D-3)(2k+N_u-N_v)+2}}\,.\label{def_s}
\end{equation}
Here 
\begin{equation}
p=(v-\Phi(\rho))\rho^{D-4}\,,\qquad q=u\rho^{-(D-2)}\,.\label{def_pq}
\end{equation}
The solution for $f^{(k)}(p,q)$ at a certain order $k$ in perturbation theory after this 2D reduction is
\begin{equation}
f^{(k)}(p,q)=\int dq' \int dp'\, G^k_m(p,q;p',q')s^{(k)}(p',q')\,,\label{f_volume}
\end{equation}
where $G^{k}_m$ is the reduced Green's function,
\begin{equation}
G^{k}_m(p,q;p',q')=-\frac{1}{4}\int_0^\infty dy\, y^{\frac{D-4}{2}-(D-3)(2k+N_u-N_v)}I_m^{D,0}(x_\star)\,,
\end{equation}
and $x_\star$ now reads
\begin{eqnarray}
x_\star&=&\frac{1+y^2-(q-q'y^{D-2})(p-p'y^{-(D-4)}-\Psi(y))}{2y}\,,\label{x_star_pq}\\
\Psi(y)&\equiv&\left\{
\begin{array}{ll}
\Phi(y)\ , &  D=4\  \vspace{2mm}\\
\displaystyle{ \Phi(y)-\frac{2}{D-4}}\ , & D>4\ \label{psi}\,
\end{array} \right. \ .
\end{eqnarray}
Thus all the quantities that we might need, at any order in perturbation theory, can be computed numerically as a (at most) two-dimensional integral. Surface terms are a particular case with a structure similar to the Green's function,
\begin{equation}
f^{(k)}_{S}(p,q)=(-1)^{D+1}k!f^{(k)}_0(1)\left(\frac{2}{q}\right)^k\int_0^\infty dy\,y^{\frac{D-4}{2}-(D-3)(k+N_u-N_v)}I_m^{D,k}(x_S)\,,\label{f_surface}
\end{equation}
where now
\begin{equation}
x_S\equiv x_\star(p'=0,q'=0)=\frac{1+y^2-q(p-\Psi(y))}{2y}\,.
\end{equation}
The functions $f^{(k)}_0(\rho)$ depend only on the initial conditions~\cite{Herdeiro:2011ck,Coelho:2015qaa}.  In the remainder we will focus the discussion on the construction of the characteristic coordinates so the specific form of $f^{(k)}_0(\rho)$ and $I_m^{D,n}(x_\star)$ will not be necessary.

The integration domain in Eqs.~\eqref{f_volume} and~\eqref{f_surface} is determined ultimately by considerations of causality between the observation point and the source points. This is encoded in the functions $I_m^{D,n}(x_\star)$ and in $x_\star$.  Considering an observation point $\mathcal{P}=(u,v,x^i)$ to the future of the collision, the integration point $\mathcal{P}'=(u',v',{x'}^i)$ must then be i) inside the future light cone of the collision (for the source to be non-zero), and ii) inside the past light cone of the observation point $\mathcal{P}$. We analyse these two conditions in the next sub-sections.

\subsection{The future light cone of the collision}
\label{future_cone}

In Brinkmann coordinates, the future light cone of the collision is defined by~\cite{Herdeiro:2011ck,Sampaio:2013faa,Coelho:2015qaa}
\begin{equation}
u=0\,\wedge\,v\geq\Phi(\rho)\qquad \vee \qquad u\geq0\,\wedge\,\bar v=\Phi(\bar\rho)+\frac{u\Phi'(\bar\rho)^2}{4}\,.
\end{equation}
On the $(p,q)$ plane, these two conditions define two important curves. The first defines the line where the initial data has support,
\begin{equation}
p\geq0 \qquad\wedge\qquad q=0\,,
\end{equation}
whereas the second one separates a flat region before the collision, from the curved region to the future of the collision. This second curve can be represented parametrically by
\begin{equation}
p(\zeta)=\Psi(\zeta)+\frac{\zeta-1}{\zeta^{D-3}}\qquad \wedge \qquad q(\zeta)=(\zeta-1)\zeta^{D-3}\,,
\end{equation}
where $\zeta\in[1,+\infty [$ is the parameter. This corresponds, in the original $D$-dimensional description, to optical null rays which are scattered at the collision and travel along its light cone. In this 2D reduction it becomes one single null ray (we call it ray~1 from here on).  A second solution, corresponds to the continuation of such optical rays into the curved region (see Fig.~\ref{5DspacetimeDiag}) after crossing the axis. This is similarly represented parametrically by
\begin{equation}
p(\zeta)=\Psi(\zeta)+\frac{\zeta+1}{\zeta^{D-3}}\qquad \wedge \qquad q(\zeta)=(\zeta+1)\zeta^{D-3}\,.\label{ray2_parametric}
\end{equation}
This curve is important because these optical rays cross at the axis of symmetry forming a caustic which creates a singularity in the metric perturbations and in the source (we call this ray~2). Thus, extra care needs to be taken to integrate near this region. In summary, looking at the lowest possible values for $p,q$ for each of the curves, we conclude that the lower bounds for the integration variables in Eq.~\eqref{f_volume} are
\begin{equation}
q'\geq0\,,\qquad p'\geq\left\{
\begin{array}{ll}
-\infty\ , &  D=4\  \vspace{2mm}\\
\displaystyle{-\frac{2}{D-4}}\ , & D>4\,
\end{array} \right. \ .\label{p_limits}
\end{equation}

\subsection{The past light cone of the observation point}
\label{past_cone}
The Green's function $G(u,v,x^i)$ introduced in Sect.~\ref{sec:overview} has support on the light cone of the observation point ($\chi=0$) for even $D$, whereas for odd $D$ it also has support inside it ($\chi\geq 0$). These conditions are equivalent, respectively, to $-1\leq x_\star \leq1$ and $x_\star\leq1$. This is indeed the domain where the functions $I_m^{D,0}$ are non-vanishing. The region with $x_\star>1$ is outside the past light cone of the event $(u,v,x^i)$.

To analyse the domain corresponding to the past light cone of the observation point we define curves $C_\pm(y)$ corresponding to the conditions stated above,  such that
\begin{equation}
C_\pm(y)\equiv (y\pm1)^2 y^{D-4}-(q-q' y^{D-2})((p-\Psi(y))y^{D-4}-p')\,.
\end{equation}
Then we have
\begin{equation}
C_-(y)\leq0 \Leftrightarrow x_\star-1\leq0\,,\qquad C_+(y)\geq0 \Leftrightarrow x_\star+1\geq0\,.
\end{equation}
Both curves start at a non-negative value for $y=0$, $C_\pm(0)\geq0$, and grow to infinity for large $y$. Moreover, $C_+(y)\geq C_-(y)\, \forall\, y\in\mathbb{R}_0^+$. So the domain becomes non-empty when the curve $C_-(y)$ crosses the horizontal axis. The limiting case occurs when the curve is  tangent to the $y$ axis. If the crossing exists, then the condition $C_-(y)\leq0$ gives a finite domain for the $y$ integration. The other condition, $C_+(y)\geq0$, comes into play when $C_+(y)$ starts crossing the $y$ axis, in which case the $y$ domain is broken into two.

The first condition defines the boundary of the light cone $(p',q')$ of the event $(p,q)$, whereas the second condition provides the location of the singularity of the reduced Green's function $G^k_m(p,q;p',q')$. They are the solutions to
\begin{equation}
C_\pm(y)=0 \qquad \wedge \qquad \frac{d}{dy}C_\pm(y)=0\,.
\end{equation}
For each case there are two solutions, parameterised by $y\in\mathbb{R}_0^+$ and labeled by $n=\pm1$, 
\begin{eqnarray}
p'&=&y^{D-4}\left[p-\Psi(y)-(1\pm y)\Delta_n(p,q)\right]\,,\label{eq:parametric1}\\
q'&=&\frac{1}{y^{D-2}}\left[q-\frac{1\pm y}{\Delta_n(p,q)}\right]\,,\label{eq:parametric2}
\end{eqnarray}
where
\begin{equation}
\Delta_\pm(p,q)=\frac{1\pm\sqrt{1+(2+(D-4)p)(D-2)q}}{(D-2)q} \; .
\end{equation}
If we choose the negative sign solution, then the two possibilities $\Delta_\pm$ give two characteristics going through $(p,q)$. They define the light cone of the observation point for the two dimensionally reduced problem. If we choose the $+$ solution one can check that the curve with $\Delta_+$ is inside the past light cone whereas the curve with $\Delta_-$ is inside the future light cone of $(p,q)$. The latter can be identified as curves where the Green's function has a singularity. Later we will see, in the conformal diagram, that the origin of this is an axis singularity, similar to the one for the sources.

\begin{figure}[t]
\hspace{-3mm}
\includegraphics[width=\linewidth,clip=true,trim= 0 0 0 0]{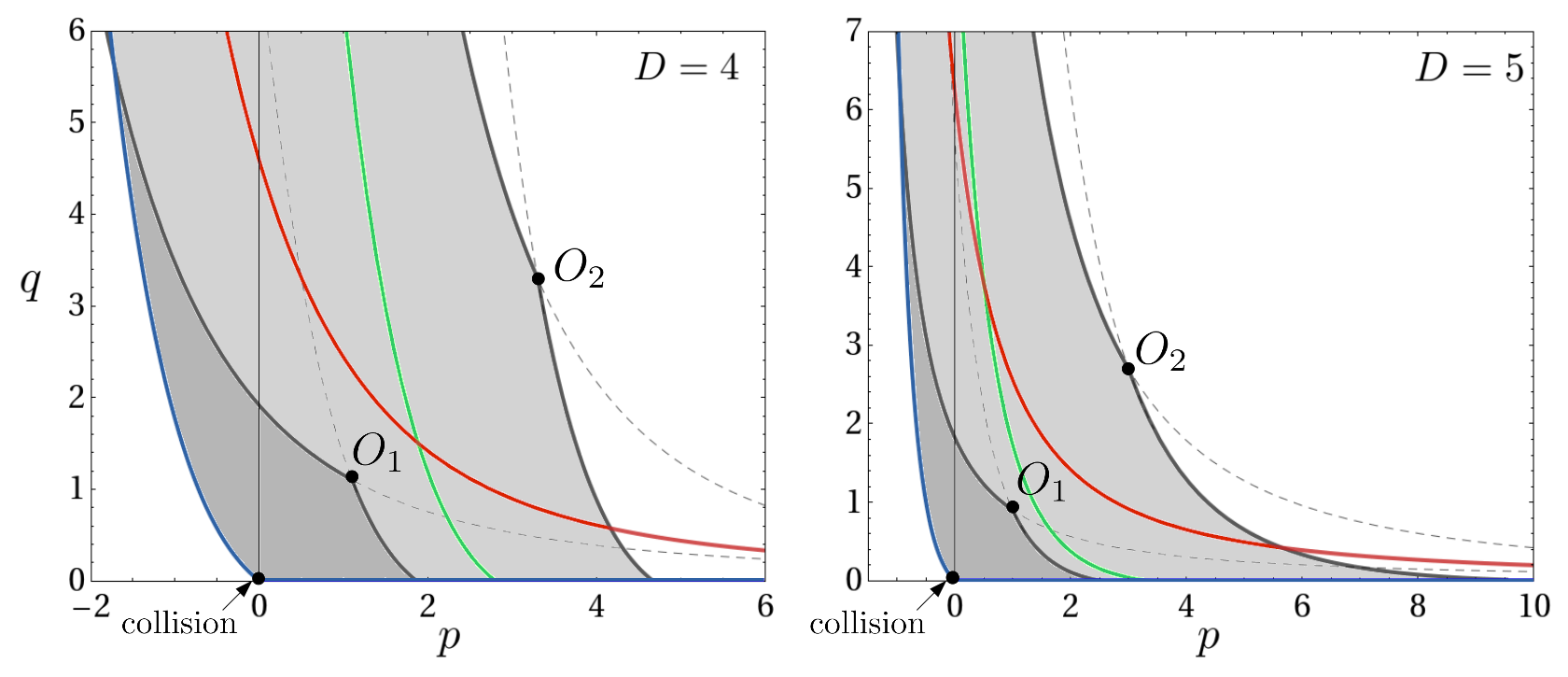}
\caption[Characteristic curves in the $(p,q)$ plane.]{\label{char_pq}Characteristic curves in the $(p,q)$ plane. We show the future light cone of the collision (blue curves), the past/future light cones for two observers $O_1$ and $O_2$ (solid/dashed grey curves), and the singularities of the source (red curve) and of the Green's function (green curve, for observer $O_2$). In $D=4$ (left), the blue curve goes to $p=-\infty$, whereas in $D>4$ (right) is starts at $p=-2/(D-4)$.}
\end{figure}
In Fig.~\ref{char_pq} we illustrate all these curves in the $(p,q)$ plane for $D=4$ and $D=5$. The collision occurs at $(p,q)=(0,0)$ and the blue curves define the future light cone of that event: the one to the right is where the initial data has support; the one to the left (ray 1) goes to $p\rightarrow-\infty$ in $D=4$ and to a constant in $D>4$ -- see also Eq.~\eqref{p_limits}. We also represent two observation points, $O_1$ and $O_2$, together with their past light cones (solid grey lines) and future light cones (dashed grey lines).  The interior of their past light cones are respectively coloured in dark grey and light grey for $O_1$ and $O_2$.\footnote{Observe, however, that the past light cone of $O_1$ is contained in the past light cone of $O_2$.} The future light cones of these events are inside the dashed grey lines to the right of the diagram. Finally, the red curve indicates the second optical ray (ray 2), where the sources are expected to be singular. The green curve shows the location of the singularity of the Green's function which is inside the past light cone of observer $O_2$.

It is also possible to write explicit equations for these curves without using a parameter. We first define a new set of coordinates
\begin{equation}
P\equiv
\left\{
\begin{array}{ll}
 p\ , &  D=4\  \vspace{2mm}\\
\displaystyle{p+\frac{2}{D-4}}\ , & D>4\ \label{def_PQ}\,
\end{array} \right. \ ,
\qquad Q\equiv (D-2)(2+(D-4)p)q\,.
\end{equation}
Then, using the parametric forms Eqs.~\eqref{eq:parametric1} and~\eqref{eq:parametric2} we solve for
\begin{equation}
y=\frac{\sqrt{1+Q}-n(D-3)}{\sqrt{1+Q'}\pm n(D-3)}\,.
\end{equation}
Inserting $y$ back in the parametric equations, we find that they can be written as
\begin{equation}
\mathcal{C}_{\pm n}(P',Q')=\mathcal{C}_{-n}(P,Q)\,,
\end{equation}
with 
\begin{equation}
\mathcal{C}_\pm(P,Q)\equiv\left\{
\begin{array}{ll}
 \dfrac{P-1}{2}+\ln\left(\dfrac{\sqrt{1+Q}\pm1}{2}\right)\pm\dfrac{1}{\sqrt{1+Q}\pm1}\ , &  D=4\ \vspace{5mm} \\
\displaystyle{\left(\dfrac{P}{\sqrt{1+Q}\pm1}\right)^{\frac{1}{D-3}}\frac{\sqrt{1+Q}\pm(D-3)}{D-2}-(D-4)^{-\frac{1}{D-3}}}\ , & D>4\ \label{L_PQ}\,
\end{array} \right. \ .
\end{equation}
Using this form we can now define the light cone $(P',Q')$ of the event $(P,Q)$ explicitly as
\begin{equation}
\mathcal{C}_\pm(P',Q')=\mathcal{C}_\pm(P,Q)\,,
\end{equation}
and the points $(P',Q')$ where the Green's function for the observation point $(P,Q)$ is singular as
\begin{equation}
\mathcal{C}_\pm(P',Q')=\mathcal{C}_\mp(P,Q)\,.
\end{equation}
Though we have obtained the equations for these characteristic curves using causality/geometric arguments, they can also be obtained by requiring that the differential operator for the problem adopts a characteristic form. This is discussed in the next section and provides a verification of this construction.

\subsection{Characteristic coordinates}
\label{characteristics}
In our initial formulation of the problem we have presented the formal solutions in terms of the coordinates $(u,v,x^i)$ which are adapted to the initial data on the characteristic surface $u=0$. After the reduction of the problem to two-dimensions, it is more natural to find new characteristic coordinates $(\xi,\eta)$ such that the principal part of the 2D reduced wave operator contains only the mixed derivative term $\partial_\xi\partial_\eta$. Then the initial data is on a characteristic line, say, of constant $\xi$.

In~\ref{app:box2D} we obtain the differential operator acting on $f^{(k)}(p,q)$, Eq.~\eqref{app:op2D}. The terms with highest derivatives come from
\begin{equation}
-4\partial_p\partial_q+\left((2+(D-4)p)\partial_p-(D-2)q\partial_q\right)^2+\ldots\,,
\end{equation}
where we ignore first derivatives. After transforming first to the coordinates $(P,Q)$ defined in Eq.~\eqref{def_PQ} we apply a generic two dimensional coordinate transformation
\begin{equation}
(P,Q)\rightarrow \left(\xi(P,Q),\eta(P,Q)\right)\,,
\end{equation}
so that the above operator takes the form
\begin{equation}
f(Z_\eta,Z_\xi)\dfrac{\partial \eta}{\partial P}\dfrac{\partial \xi}{\partial P}\dfrac{\partial^2}{\partial \eta \partial \xi}+\left(\dfrac{\partial \eta}{\partial P}\right)^2C(Z_\eta)\dfrac{\partial^2}{\partial \eta^2}+\left(\dfrac{\partial \xi}{\partial P}\right)^2C(Z_\xi)\dfrac{\partial^2}{\partial \xi^2}+\ldots\,,
\end{equation}
where $Z_X\equiv\partial_Q X/\partial_PX$. The characteristic polynomial is
\begin{equation}
C(Z) \equiv (D-4)^2P^2+4Q(Q-(D-2)(D-4))Z^2-4(D-4)P(Q+D-2)Z\,,
\end{equation}
and the explicit form of $f(Z_\eta,Z_\xi)$ in unimportant. The new coordinates $(\xi,\eta)$ are characteristics if $C(Z_\xi)=C(Z_\eta)=0$, or equivalently
\begin{eqnarray}
\dfrac{2Q(Q-(D-4)(D-2))}{Q+(D-2)(1-\sqrt{1+Q})}\dfrac{\partial \eta}{\partial Q}&=&(D-4)P\dfrac{\partial \eta}{\partial P}\,, \\
\dfrac{2Q(Q-(D-4)(D-2))}{Q+(D-2)(1+\sqrt{1+Q})}\dfrac{\partial \xi}{\partial Q}&=&(D-4)P\dfrac{\partial \xi}{\partial P}\,.
\end{eqnarray}
These allow for a solution by separation of variables and,  with a convenient choice of normalisation, a possible solution is precisely\footnote{In $D=4$ this reproduces the results of D'Eath and Payne \cite{D'Eath:1992hd,Payne} if we note that $(D-4)P\rightarrow 2$.}
\begin{equation}
\xi=\mathcal{C}_-(P,Q)\,,\qquad \eta=\mathcal{C}_+(P,Q)\,.
\end{equation}
In conclusion, we confirm that all the curves found in the previous section are indeed characteristics. Expressed in the new characteristic coordinates, the light cone events $(\xi',\eta')$ of an event $(\xi,\eta)$ are defined by
\begin{equation}
\xi'=\xi\,,\qquad \eta'=\eta\,.
\end{equation}
The events that we have identified as singularitues of the Green's function are respectively defined by $\xi'=\eta$, inside the past light cone of $(\xi,\eta)$, and $\eta'=\xi$, inside the future light cone of $(\xi,\eta)$. Observe that we have chosen the integration constants such that: i) $\eta=0$  corresponds to the left blue curve in Fig.~\ref{char_pq} going through the event $(p,q)=(0,0)$, and ii) $\xi=0$ corresponds to the second optical ray (ray 2, or red curve in Fig.~\ref{char_pq}) obtained in parametric form in Eq.~\eqref{ray2_parametric}.

From the results above it is straightforward to conclude that the ranges of the characteristic coordinates in the future of the collision are
\begin{equation}
\xi\in]-\infty,+\infty[\,,\qquad \eta\in[0,+\infty[\,,
\end{equation}
with $\xi\leq\eta$ inside the light cone. This suggests introducing compactified coordinates,
\begin{equation}
\hat\xi=\frac{\xi}{\sqrt{1+\xi^2}}\,,\qquad \hat\eta=\frac{\eta}{\sqrt{1+\eta^2}}\,,
\end{equation}
such that the integration domain becomes
\begin{equation}
\hat\xi\in[-1,1]\,,\qquad \hat\eta\in[0,1]\,.
\end{equation}
Then the volume integrals, Eq.~\eqref{f_volume}, become
\begin{equation}
f^{(k)}(\hat\xi,\hat\eta)=\int_{-1}^{\hat\xi} d\hat\xi' \int^{\hat\eta}_{\max \{0,\hat\xi'\}} d\hat\eta'\, \left|\frac{\partial(p',q')}{\partial(\hat\xi',\hat\eta')}\right|G^k_m(\hat\xi,\hat\eta;\hat\xi',\hat\eta')s^{(k)}(\hat\xi',\hat\eta')\,,\label{f_etaxi}
\end{equation}
where the Jacobian determinant is
\begin{eqnarray}
\left|\frac{\partial(p,q)}{\partial(\xi,\eta)}\right|&=&\left|\frac{\partial(p,q)}{\partial(P,Q)}\right|\times\left|\frac{\partial(P,Q)}{\partial(\xi,\eta)}\right|\times\left|\frac{\partial(\xi,\eta)}{\partial(\hat\xi,\hat\eta)}\right|\,,\\
&=&\left(\frac{D-3}{D-4}\right)^2\frac{P^{-\frac{2}{D-3}}Q^{\frac{D-2}{D-3}}}{\sqrt{1+Q}}\frac{1}{(1-\hat\eta^2)^{\frac{3}{2}}}\frac{1}{(1-\hat\xi^2)^{\frac{3}{2}}}\,.\label{jacobian}
\end{eqnarray}

\section{The Carter-Penrose diagram and higher order calculations}
\label{CP_diagram}
\begin{figure}[t]
\centering
\includegraphics[width=0.6\linewidth,clip=true,trim= 0 0 0 0]{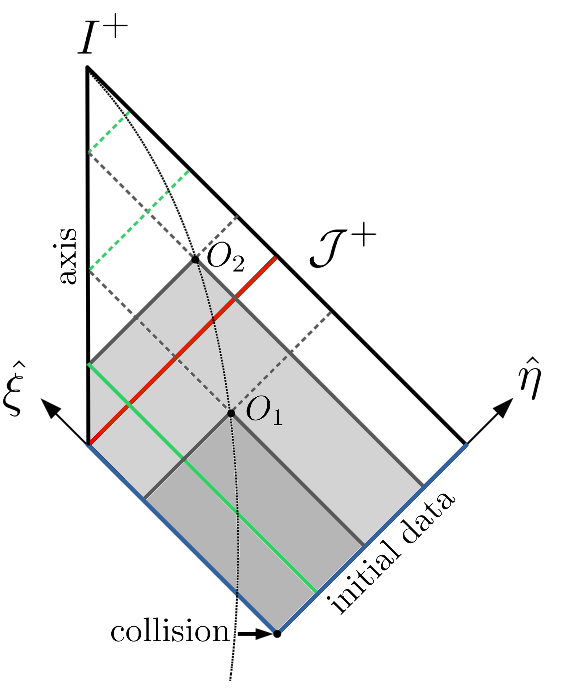}
\caption{\label{penrose}Carter-Penrose diagram for the effective background two-dimensional space-time. The colours of the lines match the corresponding curves of Fig.~\ref{char_pq}.}
\end{figure}
The compactified characteristic coordinates $(\hat\xi,\hat\eta)$ allow us to produce a conformal Carter-Penrose diagram, Fig.~\ref{penrose}, which captures, in a simple representation, the effective background two-dimensional space-time. In this diagram, the causal structure and all the important curves that we have discussed become very clear. 

The collision occurs at $(\hat\xi,\hat\eta)=(-1,0)$. From this event two light rays emerge (in blue) to define  its future light cone. At the surface $\hat\xi=-1$ the initial data is provided, and the ray $\hat\eta=0$ is the first optical ray emerging from the collision. The left vertical boundary (thick black line) is the axis, $\rho=0$, which corresponds to $\hat\xi=\hat\eta$. Future null infinity, $\mathcal{J}^+$, is located at $\hat\eta=1$ and future time-like infinity ($I^+$) at $(\hat\xi,\hat\eta)=(1,1)$. We represent, in the dotted black curve, the world line of a time-like observer. When the observer crosses the blue line (ray 1) the radiation signal begins,  and later, while crossing the red line, there is a peak in the radiation signal (corresponding to ray 2).

The light cones of two observation points $O_1$ and $O_2$, are also shown. These are solid gray lines, for the past light cones, and dashed gray lines, for the future cone. We have coloured the interior of the past light cones in dark and light gray for $O_1$ and $O_2$ respectively. Note that the singularity of the Green's function for each observation point (green solid and dashed lines) occurs for points on its light cone which cross the axis of symmetry (for past/future directed rays as seen in the solid/dashed green lines).  The solid green line, for example, is the location of such singularity in the past light cone of $O_2$. It also becomes clear that for $O_1$ ($\hat\xi<0$) this does not exist in the past light cone of $O_1$ because ray 2 (the red line) is the first ray emerging from the axis.

In this diagram, the radiation signal is extracted at $\mathcal{J}^+$ for an asymptotic observer when $r\rightarrow +\infty$ (respectively $\hat\eta\rightarrow1$). Since ultimately one is interested in extracting the radiation signal at null infinity, we now simplify all the quantities that are necessary for an asymptotic observer to perform the computation in the two-dimensional reduction. 

The characteristic coordinates are indeed the most natural for such an asymptotic analysis. In fact, if we take the limit $r\rightarrow\infty$ with $\tau$ and $\theta$ fixed, we find that
\begin{equation}
\xi\rightarrow\left\{
\begin{array}{ll} \bar{\tau}(\tau,\theta)-1\ , &  D=4\ \vspace{5mm} \\
\displaystyle{\frac{1}{(D-4)^{\frac{1}{D-3}}}\left(\frac{D-4}{D-3}\bar\tau(\tau,\theta)-1\right)}\ , & D>4\,
\end{array} \right.\,, \qquad \eta\rightarrow\frac{2(D-3)}{(D-2)^{\frac{D-2}{D-3}}}\frac{1}{\hat q}\sim O(r)\,.\,
\end{equation}
Here $\bar\tau(\tau,\theta)$ is precisely a time coordinate that we have used in the proof of the factorisation of the angular dependence of the news function in Ref.~\cite{Coelho:2014gma}. Therefore, at null infinity when $\eta\rightarrow\infty$, the natural time integration variable is indeed $\xi$. 

Now that we understand the natural coordinates to treat the problem at $\mathcal{J}^+$ we need to find the corresponding 2D reduced quantities that are necessary to compute the wave form, Eq.~\eqref{f_volume}. Since the source function $s^{(k)}(p',q')$ is already in a 2D form, it remains to analyse the Green's functions $G^k_m(p,q;p',q')$. We first observe that all the dependence on the observation point $(p,q)$ is in $x_\star$. In $(P,Q)$ coordinates, from Eq.~\eqref{x_star_pq}, we know that in $D=4$
\begin{eqnarray}
x_\star&=&\frac{4-Q(P-P'+2\ln y)}{8y}+\frac{4+Q'(P-P'+2\ln y)}{8}y\,,
\end{eqnarray}
and that in $D>4$
\begin{eqnarray}
x_\star&=&
\frac{(D-2)(D-4)-Q}{2(D-2)(D-4)y}+\frac{(D-2)(D-4)-Q'}{2(D-2)(D-4)}y\nonumber\\
&&+\frac{QP'/P}{2(D-2)(D-4)y^{D-3}}+\frac{Q'P/P'}{2(D-2)(D-4)}y^{D-3}
\end{eqnarray}
For a general observation point it is manifest, in these coordinates, that $x_\star$ is a three-dimensional quantity, depending only on $Q$, $Q'$, and $P-P'$ (in $D=4$) or $P/P'$ (in $D>4$). However, we are interested in the limit $r\rightarrow+\infty$ (equivalently $\eta\rightarrow +\infty$), i.e. for an onservation point in $\mathcal{J}^+$. When taking that limit we will loose one coordinate and the Green function will effectively depend only on two coordinates, so it becomes a 2D quantity as well. To do so, recall that $q\rightarrow0$ and $p\rightarrow\infty$ with $pq\rightarrow1$, so that
\begin{equation}
P\rightarrow\infty\,,\qquad Q\rightarrow (D-2)(D-4)\,.
\end{equation}
To avoid a divergence in the term proportional to $P$, we scale the integration variable $y$ as follows
\begin{equation}
y=\left\{
\begin{array}{ll} \dfrac{Q}{Q'}\hat y			\ , &  D=4\ \vspace{5mm} \\
\displaystyle{	\left(\frac{P'}{P}\right)^{\frac{1}{D-3}}\hat y	}\ , & D>4\,\label{y_asympt}
\end{array} \right.\,,
\end{equation}
such that, asymptotically,
\begin{equation}
x_\star\rightarrow\left\{
\begin{array}{ll} \dfrac{1}{2\hat y}\left[\hat y^2-\dfrac{Q'}{2}\left(\ln\dfrac{4\hat y}{Q'}+1-\dfrac{\Delta'}{2}\right)\right]		\ , &  D=4\ \vspace{5mm} \\
\displaystyle{\frac{1}{2\hat y^{D-3}}\left(1+\frac{Q'\hat y^{2(D-3)}}{(D-2)(D-4)}-\frac{2(D-3)\hat y^{D-4}}{(D-4)\Delta'^{\frac{1}{D-3}}}\right)}\ , & D>4\,\label{x_asympt}
\end{array} \right.\,,
\end{equation}
where
\begin{equation}
{\Delta'}^{\frac{1}{D-3}}\equiv\left\{
\begin{array}{ll} P'-P+2+\dfrac{4}{Q}-2\ln Q \qquad\,\qquad  \rightarrow P'-2\xi\,, &  D=4\ \vspace{5mm} \\
\displaystyle{     \left(\frac{P'}{P}\right)^{\frac{1}{D-3}}\left(\frac{2(D-2)(D-3)}{Q-(D-2)(D-4)}\right) \rightarrow \frac{{P'}^{\frac{1}{D-3}}}{1+(D-4)^{\frac{1}{D-3}}\xi}    }\ , & D>4\,
\end{array} \right.\,.\label{Delta_p}
\end{equation}
Now it is clear that $x_\star$ only depends on the observation and source point through $(\Delta',Q')$. Thus, factoring out the scaling factor in Eq.~\eqref{y_asympt}, the asymptotic Green's function becomes effectively 2D:
\begin{equation}
G_m^k(p,q;p',q')\rightarrow
\left\{
\begin{array}{ll}  \left(\dfrac{Q}{Q'}\right)^{1-(2k+N_u-N_v)}G_m^k(\Delta',Q')	, &  D=4\ \vspace{5mm} \\
\displaystyle{    \left(\dfrac{P'}{P}\right)^{\frac{1}{2}\frac{D-2}{D-3}-(2k+N_u-N_v)}G_m^k(\Delta',Q')    }\ , & D>4\,\label{GreenF_2D}
\end{array} \right.\,.
\end{equation}
Observe that the special case of surface terms can be obtained by setting $q'=p'=0$ in $x_\star$ similarly to the Green's function. The scalling factor for $y$ is as in Eq.~\eqref{y_asympt} without $Q'$ or $P'$. Then, using Eq.~\eqref{f_surface} we obtain
\begin{equation}
q^k f_S^{(k)}(p,q)\rightarrow
\left\{
\begin{array}{ll}  Q^{1-(k+N_u-N_v)}f_S^{(k)}(\xi)	, &  D=4\ \vspace{5mm} \\
\displaystyle{    P^{-\frac{1}{2}\frac{D-2}{D-3}+(k+N_u-N_v)}f_S^{(k)}(\xi)    }\ , & D>4\,
\end{array} \right.\,.\label{surface_scri+}
\end{equation}
This result is useful to compute the asymptotic behaviour of the second-order source $s^{(2)}(p',q')$ near $\mathcal{J}^+$.

The asymptotic behaviour of the Green's function is important for the evaluation of the wave forms at higher orders. This determines the tail off behaviour of the metric functions and needs to be factored out for a successful evaluation of the radiated power. Using Eq.~\eqref{GreenF_2D}, noting that $Q\sim r^{-1}$ in $D=4$ and $P\sim r^{D-3}$ in $D>4$, one concludes that the power of $r$ (or, equivalently, of $\eta$) controlling the decay of the Green's funtion is
\begin{equation}
\frac{D-2}{2}-(D-3)(2k+N_u-N_v)\,.
\end{equation}
Then, finally, the natural definition of the asymptotic metric function $\hat f^{(k)}(\hat\xi)$ is the finite limit
\begin{eqnarray}
\hat f^{(k)}(\hat\xi)&\equiv&\lim_{\eta\rightarrow\infty}f^{(k)}(\hat\eta,\hat\xi)\times\left\{
\begin{array}{ll}  \left(\dfrac{Q}{4}\right)^{1-(2k+N_u-N_v)}	, &  D=4\ \vspace{5mm} \\
\displaystyle{    \left(\frac{1}{P}\right)^{\frac{1}{2}\frac{D-2}{D-3}-(2k+N_u-N_v)}   }\ , & D>4\,
\end{array} \right.\,,\nonumber\\
&=&\int_{-1}^{\hat\xi} d\hat\xi' \int^{1}_{\max \{0,\hat\xi'\}} d\hat\eta'\, G^k_m(\Delta',Q')s^{(k)}(\hat\xi',\hat\eta')J^k(\hat\xi',\hat\eta')\,,\label{f_xi}
\end{eqnarray}
where
\begin{equation}
J^k(\hat\xi',\hat\eta')\equiv\left|\frac{\partial(p',q')}{\partial(\hat\xi',\hat\eta')}\right|\times\left\{
\begin{array}{ll}  \left(\dfrac{4}{Q'}\right)^{1-(2k+N_u-N_v)}	, &  D=4\ \vspace{5mm} \\
\displaystyle{    {P'}^{\frac{1}{2}\frac{D-2}{D-3}-(2k+N_u-N_v)}   }\ , & D>4\,
\end{array} \right.\,.\label{def_J_k}
\end{equation}

These results imply that both the Green's function and the source functions can be tabulated independently on a two-dimensional domain. This has been imperative to make the double integration efficient in $D=4$ and is expected to be important in $D>4$ and at higher orders. In fact, there is another choice of coordinates which is more natural for the Green's function. These are defined
\begin{eqnarray}
\delta\xi&\equiv&\mathcal{C}_-(\Delta',Q')=\frac{\xi'-\xi}{1+(D-4)^{\frac{1}{D-3}}\xi}\,,\label{def_delta_xi}\\
\delta\eta&\equiv&\mathcal{C}_+(\Delta',Q')=\frac{\eta'-\xi}{1+(D-4)^{\frac{1}{D-3}}\xi}\,,\label{def_delta_eta}
\end{eqnarray}
together with their compactified versions,
\begin{equation}
\delta\hat\xi\equiv\frac{\delta\xi}{\sqrt{1+\delta\xi^2}}\,,\qquad \delta\hat\eta\equiv\frac{\delta\eta}{\sqrt{1+\delta\eta^2}}\,,
\end{equation}
and they are naturally given as a shift around the observation point $(\xi,\eta)$. These definitions are very similar to those of $\eta$ and $\xi$ through $\mathcal{C}_\pm(P',Q')$, with the exception that, while $P'\geq0$, $\Delta'$ can also be negative. 

To finalise, we remark on the domain for these coordinates. Given an observation time $\xi$, the domain of $(\delta\xi,\delta\eta)$ depends on $\xi$. However, in a practical application where the Green's function is tabulated, we are interested in the full domain for all values of $\xi$. In $D=4$ we obtain
\begin{equation}
\delta\hat\xi\in [-1,0]\,,\qquad \delta\hat\eta\in[-1,1]\,,\qquad \text{with}\qquad \delta\hat\xi<\delta\hat\eta\,.
\end{equation}
For $D>4$ the denominator of Eqs.~\eqref{def_delta_eta}-\eqref{def_delta_xi} changes sign at $\xi=\xi_0\equiv-(D-4)^{-\frac{1}{D-3}}$. One can show that the full domain in this case is
\begin{equation}
\delta\hat\xi\in[-1,0]\,,\qquad \delta\hat\eta\in [\hat\xi_0,1]\,,\qquad \text{with}\qquad \delta\hat\xi<\delta\hat\eta\,,
\end{equation}
together with a rectangle,
\begin{equation}
\delta\hat\xi\in[0,1]\,,\qquad \delta\hat\eta\in [-1,\hat\xi_0]\,.
\end{equation}

This completes the description of the analytic structure of this problems at all orders in perturbation theory in the two-dimensional reduction for all $D$.

\section{Conclusions}
\label{sec:conclusions}

In this article we have closed the discussion of the analytic structure of the 2D reduced problem for the collision of two $D$-dimensional Aichelburg-Sexl gravitational shock waves. 

In our previous work~\cite{Herdeiro:2011ck,Coelho:2012sya,Coelho:2012sy,Coelho:2014gma,Coelho:2014qla,Coelho:2013zs,Coelho:2014cna,Sampaio:2013faa,Coelho:2015qaa} we have extended the $D=4$ method of D'Eath and Payne to $D>4$ and proved important results at all orders. Namely, the correspondence between the order of the axis expansion and the order in perturbation theory and the calculation of all surface terms in exact analytic form~\cite{Coelho:2014gma}. However, the generalisation of the characteristic coordinates to $D>4$ was still missing, and the construction of a conformal diagram for this 2D reduction did not exist at all for any $D$. The latter greatly simplifies the interpretation of the various causal boundaries and light rays in the problem. It allows the identification of the singular points for the source functions and Green's functions at fixed values of the coordinates, as well as the location of the boundaries of the integration domain. These are crucial for the successful numerical integration to obtain the metric perturbations at any order. Finally, we have shown that both source functions and Green's functions effectively only depend on two variables each, so they can be numerically tabulated in a 2D domain, at all orders.

The results in this article, together with those in previous studies~\cite{D'Eath:1992hb,D'Eath:1992hd,D'Eath:1992qu,Herdeiro:2011ck,Coelho:2012sya,Coelho:2012sy,Coelho:2014gma,Coelho:2014qla,Coelho:2013zs,Coelho:2014cna,Sampaio:2013faa,Coelho:2015qaa}, complete the discussion of this method. The computation of any higher order correction should amount to achieve a numerically stable evaluation of the various integrals involved.

\section*{Acknowledgments}
The authors are grateful to C. Herdeiro for a fruitful collaboration on this topic and continuous support. \\
F.C. and M.S. were supported by the FCT grants SFRH/BD/60272/20
09 and SFRH/BPD/69971/2010. The work in this paper is also supported by the H2020-MSCA-RISE-2015 Grant No. StronGrHEP-690904 and by the CIDMA project UID/MAT/04106/2013

\appendix

\section{2D reduced wave operator}\label{app:box2D}
After using the axial symmetry, the differential equation associated with our problem is
\begin{equation}
\Box \left(F(u,v,\rho)X_m\right)=S(u,v,\rho)X_m\,.\label{eqFX}
\end{equation}
where $X_m$ generically denotes one of the axial tensors with one of the ranks $m=\{0,1,2\}$ respectively. We can write this as
\begin{equation}
\Delta\equiv\partial^i\partial_i=\rho^{-(D-3)}\partial_\rho(\rho^{D-3}\partial_\rho)+\rho^{-2}\Delta_{S^{D-3}}\,,
\end{equation}
where $\Delta_{S^{D-3}}$ is Laplacian on the $(D-3)$-sphere, for which  $X_m$ are eigenfunctions:
\begin{equation}
\Delta_{S^{D-3}} X_m=-m(m+D-4)X_m\,,
\end{equation}
thus the equation for $F(u,v,\rho)$ implied by Eq.~\eqref{eqFX} is
\begin{equation}
\left(-4\partial_u\partial_v+\partial^2_\rho+(D-3)\rho^{-1}\partial_\rho-m(m+D-4)\rho^{-2}\right)F(u,v,\rho)=S(u,v,\rho)\,.\label{operator_3D}
\end{equation}
Using Eq.~\eqref{def_pq}, which defines $p,q$, then
\begin{equation}
\partial_u\rightarrow\rho^{-(D-2)}\partial_q\,,\qquad \partial_v\rightarrow\rho^{D-4}\partial_p\,,\qquad \partial_\rho\rightarrow\partial_\rho-\frac{D-2}{\rho}q\partial_q+\frac{2+(D-4)p}{\rho}\partial_p\,.\label{derivatives_transformation}
\end{equation}
Inserting this in Eq.~\eqref{operator_3D}, together with Eqs.~\eqref{f_pq} and~\eqref{def_s}, we conclude that the operator acting on $f^{(k)}(p,q)$ is
\begin{eqnarray}
&&-4\partial_p\partial_q+\left((2+(D-4)p)\partial_p-(D-2)q\partial_q-(D-3)(2k+N_u+N_v)+D-4\right)\times\nonumber\\
&\times&\left((2+(D-4)p)\partial_p-(D-2)q\partial_q-(D-3)(2k+N_u+N_v)\right)-m(m+D-4)\,.\nonumber\\ \label{app:op2D}
\end{eqnarray}

\bibliographystyle{h-physrev4}
\bibliography{references.bib}   

\end{document}